\newif\iffiginc
\figincfalse  
\iffiginc
\documentstyle[prd,aps,preprint,tighten,floats,psfig]{revtex}
\else
\documentstyle[prd,aps,preprint,tighten,floats]{revtex}
\fi
\begin{document}
\draft
\preprint{UPR-645-T}
\date{\it February 1995}
\title{STATIC FOUR-DIMENSIONAL ABELIAN BLACK HOLES
IN KALUZA-KLEIN THEORY}
\author{Mirjam Cveti\v c
\thanks{E-mail address: cvetic@cvetic.hep.upenn.edu}
and Donam Youm\thanks{E-mail address: youm@cvetic.hep.upenn.edu}}
\address{ Department of Physics and Astronomy \\
          University of Pennsylvania, Philadelphia PA 19104-6396}
\maketitle
\begin{abstract}
{Static, four-dimensional (4-d) black holes (BH's)
in ($4+n$)-d Kaluza-Klein (KK) theory with Abelian isometry
and diagonal internal metric have at most one electric ($Q$) and
one magnetic ($P$) charges, which can either come from the same
$U(1)$-gauge field (corresponding to BH's in effective 5-d KK theory) or
from different ones (corresponding to BH's with $U(1)_M\times U(1)_E$
isometry of an effective 6-d KK theory).
In the latter case, explicit non-extreme solutions have the
global space-time of Schwarzschild BH's, finite
temperature, and non-zero entropy.  In the extreme (supersymmetric)
limit the singularity becomes null, the temperature saturates the upper
bound $T_H=1/4\pi\sqrt{|QP|}$, and entropy is zero.
A class of KK BH's with constrained charge configurations, exhibiting a
continuous electric-magnetic duality, are generated by global $SO(n)$
transformations on the above classes of the solutions.}
\end{abstract}
\section{Introduction}
Kaluza-Klein (KK) compactification \cite{KAL}\footnote{For a recent
review see Ref. \cite{D}.} in its original sense is
the procedure by which one unifies the pure 4-dimensional (4-d) gravity
with gauge theories through the dimensional reduction of
the higher dimensional pure gravity theories.  The basic idea is that
the isometry symmetry of the internal space manifests itself as the
gauge symmetry in 4-d after the compactification of the extra space.
In addition to gauge fields, associated with the isometry group, the
scalar fields, corresponding to the degrees of freedom of the
internal metric, arise upon compactification.

A class of interesting solutions for such an effective 4-d KK theory
constitutes configurations with a non-trivial 4-d space-time
dependence for the additional scalar fields, gauge fields,
as well as for the 4-d space-time metric
\footnote{One can view \cite{CY} this class of configurations as a subset of
black holes in the heterotic string theory \cite{Sen} compactified on a six
torus. Such configurations, therefore, constitute a
class of non-trivial solutions in low-energy 4-d string theory.}.
In particular, spherically symmetric charged configurations correspond to
charged black hole (BH) solutions with additional scalar fields
varying with the spatial radial coordinate.  We shall refer to such
configurations as KK BH's.

Examples of KK BH's in 5-d \cite{FIV,DM,GW} KK theories
with Abelian isometry have been studied in the
past.\footnote{See also Ref. \cite{GM}.}  Dobiasch and Maison \cite{DM}
developed a formalism by which one can generate the most general
stationary, spherically symmetric solutions in Abelian KK theories.
Their formalism was applied primarily to obtain the explicit
solutions for the BH's in 5-d KK theory, only.

Among a class of non-trivial configurations, those which saturate
the Bogomol'nyi bound on their energy (ADM mass \cite{ADM})
can be viewed as non-trivial vacuum configurations -- solitons --
and are thus of special interest.
When embedded in supersymmetric theories such configurations
correspond to bosonic configurations which are invariant under
(constrained) supersymmetry transformations, {\it i.e.}, they
satisfy the corresponding Killing spinor equations.  One refers
to such configurations as supersymmetric ones.

Supersymmetric KK BH's in 5-d KK theory were first discussed by Gibbons
and Perry \cite{GIBB}.  Their result was recently generalized \cite{EX}
to supersymmetric KK BH's in $(4+n)$-d Abelian KK theory, {\it i.e.},
those with the $U(1)^n$ internal isometry group.  With a diagonal internal
metric, such solutions exist if and only if the isometry group of
the internal space is broken down to the $U(1)_M \times U(1)_E$ group.

The aim of this paper is to address static 4-d charged BH solutions,
compatible with the corresponding Bogomol'nyi bound,
in $(4+n)$-d KK theory with Abelian isometry group $U(1)^n$.
Explicit solutions with diagonal internal metric
turn out to correspond to configurations with at most one electric
and one magnetic charges, which can either come from the same
$U(1)$-gauge field (corresponding to dyonic BH's in effective
5-d KK theory) or from different ones (corresponding to BH's with
the $U(1)_M \times U(1)_E$ isometry in an effective 6-d KK theory).
Their global space-time and thermal properties are explored.
A class of Abelian KK BH solutions with constrained charge configurations
is further obtained by performing global $SO(n)$ transformations on
the above solutions.

The paper is organized in the following way. In Section II we discuss
dimensional reduction of the $(4+n)$-d gravity. In Section III we
derive the constraints on charges for charged KK BH's with the diagonal
internal metric Ansatz.  In Section IV we discuss the explicit form of
the non-extreme $U(1)_M \times U(1)_E$ solutions, their global
space-time structure and thermal properties.  In Section V we discuss
the electric-magnetic duality of a more general class of the solutions
generated by global $SO(n)$ rotations on the internal metric.
Conclusions are given in Section VI.
\section{Dimensional Reduction of $(4+n)$-Dimensional Gravity}
The starting point is the pure Einstein-Poincar\' e gravity in
$(4+n)$-d:
\begin{equation}
{\cal L} = -{1 \over {2\kappa^2}} \sqrt{-g^{(4+n)}}{\cal R}^{(4+n)}\ \ ,
\label{highlag}
\end{equation}
where the Ricci scalar ${\cal R}^{(4+n)}$ and the determinant
$g^{(4+n)}$ are defined in terms of a $(4+n)$-d metric
$g^{(4+n)}_{\Lambda\Pi}$, and $\kappa$ is the $(4+n)$-d
gravitational constant.  For the notation of space-time indices, we
shall follow the convention of Ref. \cite{EX}, {\it i.e.}, the upper-
(or lower-) case letters are for those running over $(4+n)$-d (or 4-d)
space-time.  Those with tilde are reserved for the $n$ extra spatial
dimensions.  Latin (or greek) letters denote flat (or curved) indices.
We shall use the mostly positive signature convention
$(-++\cdot\cdot\cdot +)$ for the $(4+n)$-d metric.

The dimensional reduction of the higher dimensional gravity is
achieved by splitting the $(4+n)$-d metric $g^{(4+n)}_{\Lambda\Pi}$
into the following form:
\begin{equation}
g^{(4+n)}_{\Lambda \Pi} =
\left [ \matrix{{\rm e}^{-{1 \over \alpha}\varphi}g_{\lambda \pi} +
{\rm e}^{{2\varphi} \over {n\alpha}}\rho_{\tilde{\lambda}
\tilde{\pi}}A^{\tilde{\lambda}}_{\lambda} A^{\tilde{\pi}}_{\pi} &
{\rm e}^{{2\varphi} \over {n\alpha}}\rho_{\tilde{\lambda}
\tilde{\pi}}A^{\tilde{\lambda}}_{\lambda}
\cr {\rm e}^{{2 \varphi} \over {n\alpha}}\rho_{\tilde{\lambda}
\tilde{\pi}}A^{\tilde{\pi}}_{\pi} & {\rm e}^{{2\varphi} \over
{n\alpha}}\rho_{\tilde{\lambda} \tilde{\pi}}} \right ]\ \ ,
\label{ansatz}
\end{equation}
where $\rho_{\tilde{\lambda} \tilde{\pi}}$ is the unimodular part,
{\it i.e.}, ${\hbox{det}}\rho_{\tilde{\lambda} \tilde{\pi}}=1$,
of the internal metric, $\varphi$ is the determinant of the
internal metric, which we will refer to as dilaton, and
$\alpha = \sqrt{{n+2}\over n}$.

The effective theory in 4-d is then obtained by imposing ``the right
invariance'' \cite{CHB} of the $(4+n)$-d metric $g_{\Lambda\Pi}$
under the action of an isometry of the internal space.  This requirement
fixes the dependence of the metric components on the internal coordinates.
It turns out that the internal coordinate dependence of the transformation
laws (under the general coordinate transformations) of the fields in
(\ref{ansatz}) factors out, and $(4+n)$-d Einstein Lagrangian density
(\ref{highlag}) becomes independent of the internal coordinates.  Then,
4-d effective action after a trivial integration over the internal space
assumes the form (see for example Eq.(8) of Ref. \cite{CHB}):
\begin{eqnarray}
{\cal L}= -{1 \over 2}\sqrt{-g}[{\cal R} +
{\rm e}^{-\alpha\varphi}{\cal R}_K + {1 \over 4}{\rm e}^{\alpha\varphi}
\rho_{ij}F^i_{\mu \nu}F^{j \mu \nu}
+ {1 \over 2}\partial_{\mu}\varphi \partial^{\mu}\varphi \nonumber \\
+ {1 \over 4}\rho^{ij}\rho^{k\ell}
(D_{\mu}\rho_{ik})(D^{\mu}\rho_{j\ell})
+ \chi (\det\rho_{ij} - 1)]\ \ ,
\label{efflag}
\end{eqnarray}
where the Ricci scalar ${\cal R}_K$ is defined in terms of the unimodular
part $\rho_{ij}$ $(i,j=1,2,...,n)$
\footnote{From now on, for the simplicity of
notation, we shall denote the internal space
indices as $i\equiv \tilde{\alpha} - 3, etc.$ }
of the internal metric, $F^i_{\mu \nu} \equiv \partial_{\mu}
A^i_{\nu} - \partial_{\nu}A^i_{\mu} - gf^i_{jk} A^j_{\mu} A^k_{\nu}$,
where $f^i_{jk}$ is the structure constant for the internal isometry
group and $g$ is the gauge coupling constant of the isometry group,
is the field strength of the gauge field $A^i_{\mu}$,
$D_{\mu} \rho_{ij} = \partial_{\mu}\rho_{ij} - f^\ell_{kj}A^k_{\mu}
\rho_{i \ell}$ is the corresponding gauge covariant derivative
and $\chi$ is the Lagrangian multiplier.
The 4-d gravitational constant $\kappa_4$ has been set equal to 1.
When the isometry group is Abelian, {\it i.e.}, one compactifies on
$n$-torus $T^n$, the structure constant $f^i_{jk}$ vanishes.  In this
case, the gauge covariant derivatives in (\ref{efflag}) become ordinary
ones and the Ricci scalar ${\cal R}_K$, which in general describes
the self-interactions among scalar fields, vanishes.
\section{Diagonal Internal Metric Ansatz and Constraints on Solutions}
We shall first address the spherically symmetric configurations
with the diagonal internal metric Ansatz.  Then, the unimodular
part of the internal metric is of the form:
\begin{equation}
\rho_{ij} = {\rm diag}(\rho_1 ,..., \rho_{n-1},
\prod^{n-1}_{k=1} \rho^{-1}_k )\ \ ,
\label{diag}
\end{equation}
the spherically symmetric {\it Ansatz} for the 4-d space-time metric
is of the form:
\begin{equation}
{\rm d} s^2 = g_{\mu\nu} {\rm d}x^{\mu} {\rm d} x^{\nu} =
-\lambda (r) {\rm d} t^2 + \lambda^{-1} (r) {\rm d} r^2
+ R(r) ( {\rm d} \theta^2 + \sin^2 \theta {\rm d} \phi^2 )
\label{sph}
\end{equation}
and the scalar fields $\varphi$ and $\rho_i$ depend on the radial
coordinate $r$, only.  The electromagnetic vector potentials take the form:
\begin{equation}
A^i_{\phi} = P_i \cos \theta \ \ \ \ \ \ \ \ \ \ \ \
A^i_t = \psi_i (r) \ \ ,
\label{vector}
\end{equation}
where $E_i (r) = -\partial_r \psi_i (r) = {{\tilde{Q}_i} \over
{R {\rm e}^{\alpha \varphi} \rho_i}}$
\footnote{The physical Electric charge $Q_i$ defined in terms of the
asymptotic behavior $E_i \sim {Q_i \over r^2}$ ($r \rightarrow \infty$)
of the electric field is related to $\tilde{Q}_i$ through
$\tilde{Q}_i = {\rm e}^{\alpha\varphi_{\infty}}\rho_{i \infty} Q_i$. Here,
$\varphi_\infty$ and $\rho_\infty$ are the constant values of the
corresponding scalar fields at $r\rightarrow \infty$.}.

The {\it Ansatz} with all the off-diagonal components of the internal
metric turned off has to be consistent with the equations of motion.
In fact, such a consistency restricts the allowed charge
configurations.  Namely, the Euler-Lagrange equations for the components
$\rho_{ij}$ of the unimodular part of the metric is given by:
\begin{equation}
{1 \over 2}{\rm e}^{\alpha\varphi(r)}[R(r)E_i(r) E_j(r) -
R^{-1}(r)P_i P_j ] + \chi R(r)\rho^{ij}(r) =
{1 \over 2}{{\rm d} \over {{\rm d}r}}
[\lambda(r)R(r){{{\rm d}\rho^{ij}(r)}\over {{\rm d}r}}]
\ \ ,
\label{rhoij}
\end{equation}
which implies that for the diagonal metric {\it Ansatz} (\ref{diag}) the
following constraints have to be satisfied:
\begin{equation}
Q_iQ_j-e^{2\alpha\varphi}\rho_i\rho_j\, P_iP_j=0\ \ \
i\neq j\ \ .
\label{ij}
\end{equation}
Eqs. (\ref{ij}) can be satisfied if and only if
\footnote{A more general constraint $Q_iQ_j-c_ic_jP_iP_j=0$ with
$c_i\equiv\rho_ie^{\alpha\varphi}$ being constant would imply the equation
of motion for $\rho_ie^{\alpha\varphi}$ with $Q_i=P_i=0$.  Thus, the
constraint  $Q_iQ_j-c_ic_jP_iP_j=0$ in turn reduces to the subset of
constraints (\ref{chcon}).}
\begin{equation}
Q_i Q_j = 0\ \ \ \ {\rm and}\ \ \ \ P_i P_j = 0 ,\ \ \ \ i \neq j \ \ .
\label{chcon}
\end{equation}
Constraints (\ref{chcon}) imply that the same type of charge,
{\it i.e.}, either electric or magnetic one, can appear in at most
one gauge field.  Consequently, the internal isometry group $U(1)^n$
is broken down to at most $U(1)\times U(1)$, with one electric and
one magnetic charges, only.

When, say, the first ($n-2$) gauge fields are turned off the first
($n-2$) components of the diagonal internal metric become constant
(${\rm e}^{{2\varphi}\over {n\alpha}}\rho_i = const.$
($i=1,...n-2$)).  Thus, the KK BH's are those of effective 6-d KK:
\begin{eqnarray}
{\tilde {\cal L}} = -{1\over 2}\sqrt{-g}
[{\cal R} + {1\over 2} \partial_{\mu}\Phi \partial^{\mu}\Phi +
{1\over 2} \partial_{\mu}\chi_{n-1}\partial^{\mu}\chi_{n-1} +
{1\over 2} \partial_{\mu}\chi_n \partial^{\mu}\chi_n  \nonumber \\
+ {1\over 4}{\rm e}^{\sqrt{2}(\Phi + \chi_{n-1})}F^{n-1}_{\mu\nu}
F^{n-1 \ \mu\nu} + {1\over 4}{\rm e}^{\sqrt{2}(\Phi +\chi_n)}
F^n_{\mu\nu} F^{n \ \mu\nu}]\ \ ,
\label{effec}
\end{eqnarray}
where $\Phi \equiv {\sqrt{2} \over \alpha}\varphi$ and
$\chi_i \equiv {1 \over \sqrt{2}}[\ln\rho_i + {{2-n}\over {n\alpha}}
\varphi ]$ $(i = n-1, n)$ with the constraint $\chi_{n-1} + \chi_n =
const$.

In general, one can, therefore, have the following qualitatively
different classes of configurations:
\begin{itemize}
\item  $Q_{n-1}=P_{n-1}=Q_n=P_n=0$, which corresponds to the ordinary 4-d
Schwarzschild BH's.
\item  Say, $Q_n = P_n =0$, which corresponds to KK BH's in effective
5-d KK theory.
\item Say, $Q_{n-1}=P_n=0$, which corresponds to a class of
KK BH's in effective 6-d KK theory, where electric and magnetic
charges arise from different $U(1)$ groups.
\end{itemize}
Schwarzschild BH's are well understood, and BH's in 5-d KK theory have
been extensively studied in Refs. \cite{FIV,DM,GW}.
We, therefore, concentrate on the study of the last class of solutions,
which correspond to the non-extremal generalization of $U(1)_M \times
U(1)_E$ supersymmetric solution, studied in Ref. \cite{EX}.
\section{Non-extreme $U(1)_M \times U(1)_E$ Solutions}
If only a pair of gauge fields are non-zero, the Lagrangian density
(\ref{efflag}) becomes that of effective 6-d KK theory \cite{EX}.
Without loss of generality, one assumes that the first
[second] gauge field to be magnetic [electric].  The Einstein's and
Euler-Lagrange equations can be cast in following form:
\begin{equation}
(\lambda R)^{\prime \prime} = 2
\label{beqn}
\end{equation}
\begin{equation}
2(\lambda^{\prime}R)^{\prime} = R^{-1}{\rm e}^{-\sqrt{2}\Phi}
\rho \tilde{Q}^2 +R^{-1}{\rm e}^{\sqrt{2}\Phi}\rho P^2
\label{aeqn}
\end{equation}
\begin{equation}
{1 \over \sqrt{2}}
(\lambda R \Phi^{\prime})^{\prime} + (\lambda R \rho^{-1}
\rho^{\prime})^{\prime}=R^{-1}{\rm e}^{\sqrt{2}\Phi}\rho P^2
\label{ceqn}
\end{equation}
\begin{equation}
-{1 \over \sqrt{2}}(\lambda R \Phi^{\prime})^{\prime} +
(\lambda R \rho^{-1} \rho^{\prime})^{\prime}=R^{-1}
{\rm e}^{-\sqrt{2} \Phi}\rho \tilde{Q}^2 \ \ ,
\label{deqn}
\end{equation}
where $\rho \equiv{\rm e}^{\sqrt{2} \chi_{n-1}}$ and the prime denotes
differentiation with respect to $r$. Recall, ${\tilde Q} =
{\rm e}^{\sqrt{2}\Phi_{\infty}} \rho_{\infty}Q$.
Eqs.(\ref{aeqn})-(\ref{deqn}) exhibit manifest electric-magnetic
duality symmetry:
$P \leftrightarrow \tilde{Q}$ and $\Phi \rightarrow -\Phi$.

Equation (\ref{beqn}) implies
$\lambda R = (r - r_+ )(r - r_+ + 2\beta )$, while
the resultant equation obtained by subtracting Eqs. (\ref{ceqn}) and
(\ref{deqn}) from Eq. (\ref{aeqn}) yields
$\lambda = {\rho \over \rho_{\infty}}\left({{r-r_+ }
\over {r-r_+ +2\beta}} \right )$.
Here $r_+$ is defined to be the outermost horizon and $\beta > 0$
is the non-extremality parameter
\footnote{Note, the role of the non-extremality parameter $\beta$
is very similar to the one used in describing non-extreme
supergravity walls \cite{CS,NON}.}.
Substitution of these relations into Eqs. (\ref{ceqn}) and (\ref{deqn}),
yields the following ordinary differential equations
\footnote{These equations are reminiscent of equations for Toda molecule.}:
\begin{eqnarray}
2{\bf P}^2{{{\rm e}^X} \over {(r-r_{+})^2}} =
[(r-r_{+})(r-r_{+}+2\beta)X^{\prime}]^{\prime}\ \ \  \nonumber \\
2{\bf Q}^2{{{\rm e}^Y} \over {(r-r_{+})^2}} =
[(r-r_{+})(r-r_{+}+2\beta)Y^{\prime}]^{\prime} \ \ ,
\label{toda}
\end{eqnarray}
where $X \equiv \sqrt{2}\Phi + 2\ln \lambda - \sqrt{2}\Phi_{\infty}$
and $Y \equiv -\sqrt{2}\Phi + 2\ln \lambda + \sqrt{2} \Phi_{\infty}$.
Here, ${\bf P}$ and ${\bf Q}$ are the ``screened'' \cite{EX} magnetic
and electric monopole charges, {\it i.e.}, ${\bf P} \equiv
{\rm e}^{{1 \over \sqrt{2}}\Phi_{\infty}} \rho^{1\over 2}_{\infty}P$ and
${\bf Q} \equiv {\rm e}^{{1 \over \sqrt{2}}\Phi_{\infty}}
\rho^{-{1\over 2}}_{\infty}Q$.  The two equations in (\ref{toda})
can be solved explicity, yielding the following explicit form of
the solutions with regular horizons:
\begin{equation}
\lambda ={{r-r_+} \over {(r-r_+ +\hat{P})^{1/2}(r-r_+ + \hat{Q})^{1/2}}}
\label{asol}
\end{equation}
\begin{equation}
R = r^2 (1-{{r_+ - 2\beta}\over r})(1-{{r_+ - \hat{P}}\over r})^{1 \over 2}
(1-{{r_+ - \hat{Q}} \over r})^{1 \over 2}
\label{bsol}
\end{equation}
\begin{equation}
{\rm e}^{\sqrt{2}(\Phi - \Phi_{\infty})} =
{{r - r_+ + \hat{Q}} \over {r - r_+ + \hat{P}}}
\label{csol}
\end{equation}
\begin{equation}
\rho = \rho_\infty {{r -r_+ +2\beta} \over {(r - r_+ + \hat{P})^{1/2}
(r - r_+ + \hat{Q})^{1/2}}}  \ \ ,
\label{dsol}
\end{equation}
where $r_+ = \beta + {{|{\bf P}| \sqrt{{\bf P}^2 + \beta^2} -
|{\bf Q}| \sqrt{{\bf Q}^2 + \beta^2}} \over {|{\bf P}| - |{\bf Q}}|}$,
$\hat{P} = \beta + \sqrt{{\bf P}^2 + \beta^2}$ and $\hat{Q} = \beta +
\sqrt{{\bf Q}^2 + \beta^2}$, while the ADM mass of the configurations
is of the form:
\begin{equation}
M = 2\beta + \sqrt{{\bf P}^2 + \beta^2} + \sqrt{{\bf Q}^2 + \beta^2}
\ \ .\label{mass}
\end{equation}
In the limit $\beta \rightarrow 0$, the above expressions
reduce to those with $r_+=r_H=|{\bf P}|+|{\bf Q}|$, $\hat{P}=|{\bf P}|$,
$\hat{Q}=|{\bf Q}|$ and the ADM mass saturates the Bogomol'nyi bound
$M_{ext}=|{\bf P}|+|{\bf Q}|$. These are supersymmetric
solutions of Ref. \cite{EX}.

Now, we shall discuss the global space-time structure and
thermal properties of the solution.  We first describe the
singularity structure of 4-d space-time defined by the metric
coefficients (\ref{asol}) and (\ref{bsol}).  For the non-extreme
solutions, {\it i.e.}, $\beta > 0$, there is a space-like singularity
at $r = r_+ - 2\beta$ which is hidden behind a horizon at
$r =r_+ $.  The global space-time of the non-extreme solutions
is that of the Schwarzschild BH (see Fig. 1).
In the supersymmetric (extreme) limit ($\beta\rightarrow 0$)
the singularity becomes null, {\it i.e.}, it coincides with the horizon
(see Fig. 2a).  In the extreme limit with either $Q$ or $P$ zero,
{\it i.e.}, supersymmetric 5-d KK BH \cite{GIBB},
the singularity becomes naked (see Fig. 2b).
\begin{figure}[p]
\vfill
\iffiginc
\hfill\psfig{figure=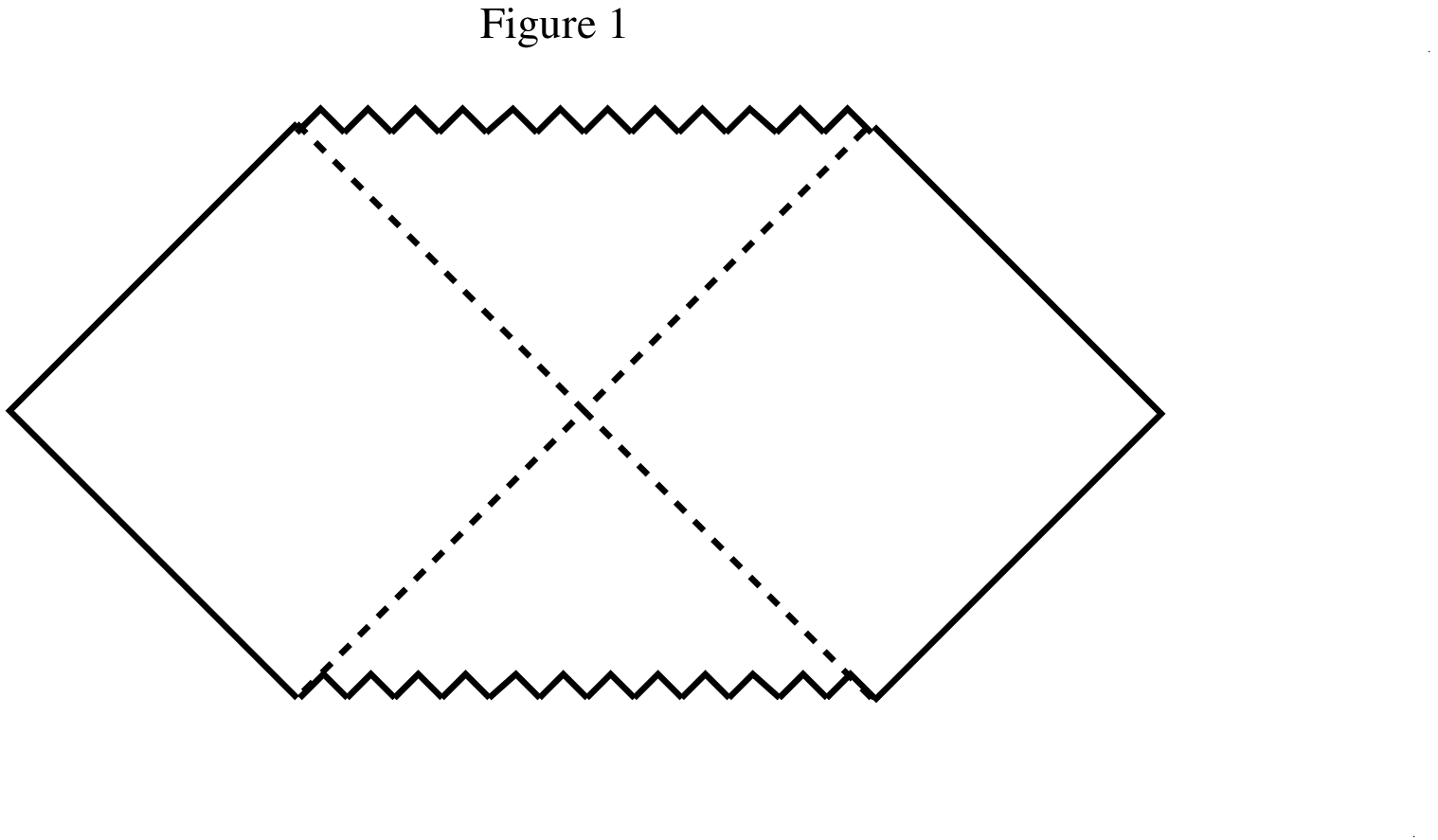}\hfill
\fi
\caption{The Penrose diagram (in the ($r,t$) plane) for non-extreme
$U(1)_M \times U(1)_E$ 4-d Kaluza-Klein black holes.
The space-like singularity (jagged line) at $r=r_+ - 2\beta$
($\beta$ is the non-extremality parameter) is hidden behind the
horizon (dashed line) at $r=r_+$.}
\label{Fig. 1}
\end{figure}
\begin{figure}
\vfill
\iffiginc
\psfig{figure=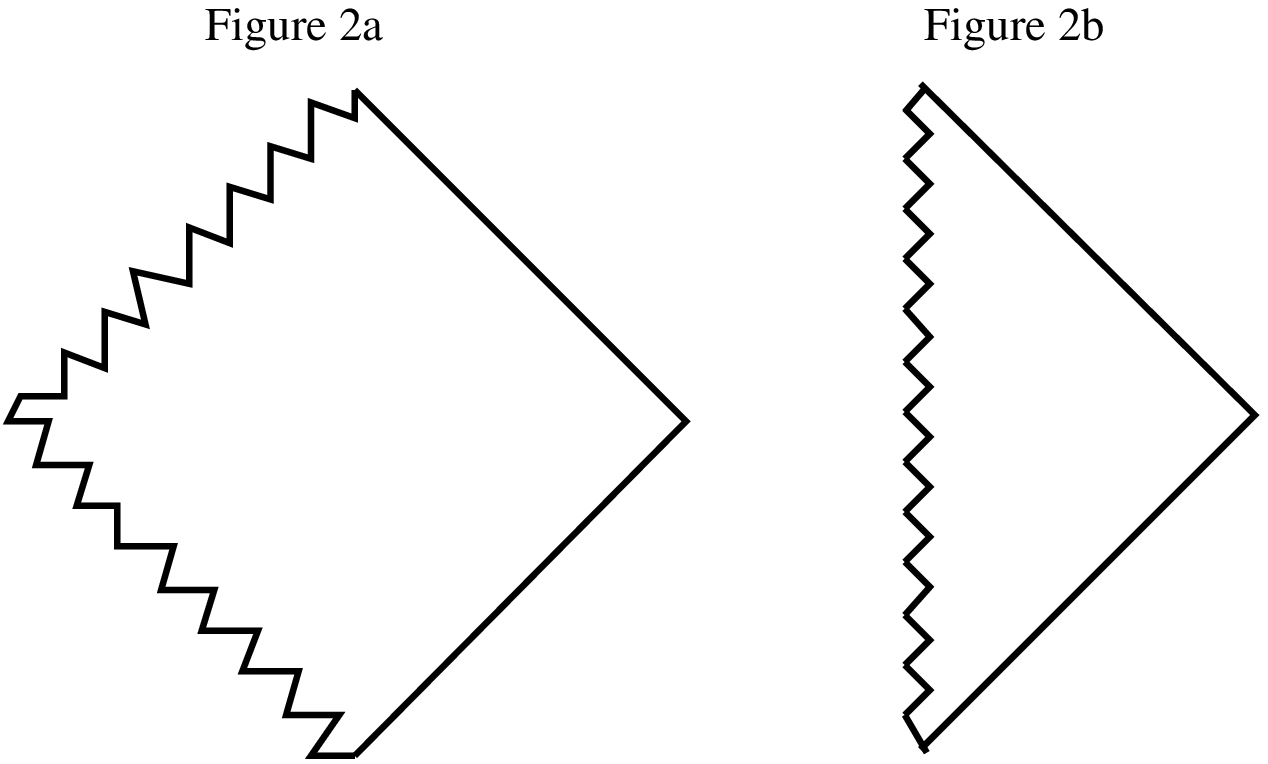}\hfill
\fi
\caption{The Penrose diagram (in the ($r,t$) plane) for the
supersymmetric (extreme) $U(1)_M \times U(1)_E$ 4-d Kaluza-Klein black
holes, {\it i.e.}, those with {\it both} $Q$ and $P$ charges
non-zero, is given in fig. 2a.  The Penrose diagram for the
supersymmetric $U(1)_E$ (or $U(1)_M$) 4-d Kaluza-Klein black
holes,{\it i.e.}, those with $P$ or $Q$ charge nonzero,
is given in fig. 2b.  Note a null singularity (jagged line) in the
former case, and a naked singularity in the latter one.}
\label{Figure 2}
\end{figure}

The Hawking temperature\ cite{HAW}, which can be calculated by
identifying the inverse of the imaginary time period \cite{GH} of a
functional path integral, turns out to be
$T_H = {{|\lambda^{\prime}(r_+ )|} \over {4\pi}}$.
With the explicit solution (\ref{asol}) for $\lambda$ one obtains:
\begin{equation}
T_H ={1 \over {4\pi [\beta + ({\bf Q}^2 + \beta^2 )^{1 \over 2}]^{1\over 2}
[\beta +({\bf P}^2 +\beta^2 )^{1\over 2}]^{1\over 2}}} \ \ .
\label{temp}
\end{equation}
As $\beta$ decreases $T_H$ increases and reaches the upper bound
$T_{H\, ext}=1/{4\pi|{\bf P Q}|}$ in the extreme limit
$\beta = 0$. In the extreme limit, when either
${\bf Q}$ or ${\bf P}$ becomes zero $T_H$ is infinite.

The entropy $S$ of the system, determined as $S = {1 \over 4}\times$
(the surface area of the event horizon) \cite{BEK}, is of the
following form:
\begin{equation}
S = 2\pi \beta [\beta +({\bf Q}^2 +\beta^2)^{1\over 2} ]^{1\over 2}
[\beta +({\bf P}^2 +\beta^2 )^{1\over 2}]^{1\over 2}\ \ .
\label{ent}
\end{equation}
where we have used the explicit solution (\ref{bsol}) for $R(r)$.
In the extreme limit the entropy becomes zero.
\section{Electric-Magnetic Duality Transformations}
In addition to the 4-d general coordinate invariance and gauge symmetry,
which arise from the $(4+n)$-d general coordinate invariance of the
$(4+n)$-d Einstein action (\ref{highlag}), the Lagrangian density
(\ref{efflag}) has a global $SO(n)$ invariance:
\begin{equation}
\rho_{ij} \rightarrow U_{ik} \rho_{k\ell} (U^{T})_{\ell j} \ \ \ \ \ \ \
A^i_{\mu} \rightarrow U_{ij} A^j_{\mu}\ \ , \label{son}
\end{equation}
where $U$ is an $SO(n)$ rotation matrix.
One convenient parameterization of $U$ is in terms of
successive rotations in all the possible planes in $\Re^n$:
\begin{equation}
U = \hat{U}_{12} \cdot \cdot \cdot \hat{U}_{1n}\hat{U}_{23}
\cdot \cdot \cdot \hat{U}_{2n} \cdot \cdot \cdot \hat{U}_{i, i+1}
\cdot \cdot \cdot \hat{U}_{in} \cdot \cdot \cdot \hat{U}_{n-1, n} \ \ ,
\label{repre}
\end{equation}
where $\hat{U}_{k\ell}$ is the rotation matrix in the $(k,\ell)$-plane
with a rotational angle $\theta_{k\ell}$.

The two types of solutions are obtained by performing
the $SO(n)$ transformation on BH solutions of the effective 5-d and
6-d KK theories, respectively.  Since the 4-d space-time metric and
the dilaton field are not affected by the $SO(n)$ transformations,
the global space-time and the thermal properties in each class of the
solutions remain the same.
The above transformations also generate non-diagonal
internal metric coefficients; the ${{n(n-1)}/2}$ degrees of freedom
associated with the rotational angles of the $SO(n)$ matrix
generate the $n(n-1)/2$ off-diagonal internal metric components.

On the other hand, the two classes of the solutions correspond to
KK BH's with constrained charge configurations, and therefore do not
constitute the most general set of static 4-d Abelian KK BH's solutions.
Namely, the subset of rotational matrices, which
transform the charge configuration of the
$U(1)_M\times U(1)_E$ solution to a new type of charge configurations,
corresponds to the coset space $SO(n)/SO(n-2)$. This subset of
transformations therefore provides $(2n-3)$ additional degrees of
freedom for the (electric and magnetic) charge configuration.
Thus, the resultant number of degrees of freedom for the charge
configuration is $2n-1$.  In fact, $2n$ ($n$-electric and $n$-magnetic)
charges $(\vec{Q},\vec{P})$ satisfy the constraint
$\vec{Q} \cdot \vec{P} = 0$.  Similarly, for solutions generated by
the $SO(n)/SO(n-1)$-transformations on the charge configuration of
the effective 5-d KK solution, the resultant number of charge degrees
of freedom is $n+1$.

In the following we shall concentrate on the electric-magnetic duality of
solutions generated by $SO(n)$ transformations on BH's of the effective
6-d KK theory, since they involve less trivial transformations than those
acting on BH's of the effective 5-d KK theory.  The unimodular part of
the internal metric and the set of $n$ gauge fields transform in
the following way:
\begin{equation}
\rho_{ij}^\prime = \rho_k {U}_{ik} {U}_{jk} \ \ \ \
A^{\prime i}_\phi= U_{i \,(n-1)} P  \cos \theta \ \ \ \
A^{\prime i}_t  = U_{in}\psi (r)  \ \ ,
\label{nondiag}
\end{equation}
where the electric and magnetic charges of new gauge field
$A^{\prime i}_\mu$ are, therefore, given by $\tilde{Q}^{\prime i} =
U_{in}\tilde{Q} $ and $P^{\prime i} = U_{i\,(n-1)}P $ ($i=1,...,n$).
The expression for 4-d metric components in the new configuration
can be obtained from (\ref{asol}) - (\ref{bsol}) by replacing
$\tilde{Q}$ and $P$ by $\sqrt{\sum^n_{i=1} (\tilde{Q}^{\prime i})^2}$ and
$\sqrt{\sum^n_{i=1}( P^{\prime i})^2}$, respectively.

The $SO(n)$ transformations generate the continuous
electric-magnetic duality transformations which rotate the
$U(1)_M \times U(1)_E$ configuration, {\it i.e.}, $P^i = \delta_{n-1}^i P$
and $\tilde{Q}^i = \delta_n^i \tilde{Q}$, to general charge
configurations $P^{\prime i} = U_{i\,(n-1)}P $ and $\tilde{Q}^{\prime i} =
U_{in}\tilde{Q}$, with the constraint $\sum^n_{i=1}P^{\prime
i}Q^{\prime i}=0$. This constraint is a consequence of
$\sum^n_{i=1} U_{i n}U_{i\, (n-1)} = 0$.

A subset of $SO(n)$ transformations corresponding to the
rotation in the $(n-1,n)$-plane, {\it i.e.}, $U=\hat U_{(n-1)\, n}$
(see Eq. (\ref{repre})), mixes the monopole charges in the
$(n-1)$-$th$ and $n$-$th$ gauge fields and induces the corresponding
off-diagonal terms in the internal metric of the effective 6-d KK
theory.  In particular, the discrete change of the rotation angle
$\theta_{(n-1)\, n}$ from 0 to $\pi \over 2$ corresponds to a discrete
electric-magnetic duality transformation, which interchanges the magnetic
and electric monopole charges in the $(n-1)$-$th$ and $n$-$th$
gauge fields \cite{EX}.

\section{Conclusions}
In this paper, we studied a class of static, spherically
symmetric solutions in $(4+n)$-d KK theory with Abelian isometry
($U(1)^n$).  In particular, for a diagonal internal metric {\it Ansatz},
the consistency of the equations of motion imposes strong constraints
on the possible charge configurations of such solutions; BH's
exist only for configurations with at most one non-zero electric
and one non-zero magnetic charges.  The case of electric and magnetic
charges coming from the same gauge field corresponds to BH's
in effective 5-d KK theory.  Configurations with electric and magnetic
charges arising from different $U(1)$ gauge factors, {\it i.e.},
the isometry group is $U(1)_M \times U(1)_E$, correspond to BH's
in effective 6-d KK theory.

Non-extreme BH's with $U(1)_M\times U(1)_E$ symmetry, which are compatible
with the corresponding Bogomol'nyi bound, are parameterized in terms of
the non-extremality parameter $\beta>0$.  They have a global
space-time structure of Schwarzschield BH's, with the temperature
$T_H$ [entropy $S$] increasing [decreasing] as $\beta$ decreases.
In the extreme limit $\beta \to 0$, the solutions correspond to
the supersymmetric BH's \cite{EX}.  In this limit, the corresponding
Bogomol'nyi bound for the ADM mass is saturated, $T_H$ [$S$] reaches
the upper [lower] limit $T_{H\, ext}=1/(4\pi\sqrt{|PQ|})$ [$S=0$],
and the space-like singularity becomes null.
Notably, the extreme limit is reached {\it smoothly}.

A class of solutions with non-diagonal internal metrics and
general charge distributions among gauge fields are obtained by
applying the global $SO(n)$ transformations on the solutions with
diagonal internal metric.  These solutions correspond to a subset of
static 4-d Abelian KK BH's with constrained charge configurations.
Since $SO(n)$ transformations act on the unimodular part of the
internal metric and the gauge fields, only, the 4-d global space-time
and the thermal properties of the general class of the 4-d abelian KK
BH's are the same as those of the corresponding 4-d abelian KK BH's
with the diagonal internal metric.
This class of solutions exhibits continuous electric-magnetic
duality symmetry parameterized by rotational angles of $SO(n)$ rotations.
\acknowledgments
The work is supported by U.S. DOE Grant No. DOE-EY-76-02-3071,
and the NATO collaborative research grant CGR
940870.  We would like to acknowledge useful discussions with E. Kiritsis,
C. Kounnas, J. Russo and D. Waldram.

\end{document}